# A Problem with Bell-type Inequalities, the Origin of the Quantum Nonlocality, and a Full/Empty Waves Model for Entanglements

*by Sofia Wechsler*

**Abstract**

Whether the quantum mechanics (QM) is nonlocal is an issue disputed for a long time. The violation of the Bell-type inequalities was considered as proving this nonlocality. However, these inequalities are constructed on a class of local hidden variables, which obey the *calculus with positive probabilities*. Such a calculus is rather suitable for billiard balls, while the QM deals with wave-packets of complex amplitudes. There is no wonder that a calculus with positive numbers doesn't match a calculus with complex numbers.

The present text describes a different model of hidden variables for entanglements, model that reproduces the quantum predictions in different experiments, and also explains why the QM is nonlocal. The model deals with *waves*, some of them *full* and the others *empty*, and the hidden variables mark which waves are full.

The basic physical concept with which the model operates is *joint amplitudes* of probability, and not probabilities. The latter are a secondary concept, the probability of a combination of results being equal to the absolute square of *sum* of all the contributing joint amplitudes.

Thus the nonlocality appears: a) a joint amplitude ignores distance, it handles distant particles as if they were one single particle at one single place, b) joint amplitudes are complex numbers and the sum of several contributions may vanish, blocking the respective combination of wave-packets and therefore of results.

Although showing the success of the model, this text *does not advocate for full/empty waves*. It is shown that this hypothesis works only as long as one doesn't consider moving observers, and doesn't compare their conclusions. The real purpose here is to point to *a severe impasse*: assuming the existence of a preferred frame contradicts the theory of relativity, while refuting the full/empty waves idea one runs into other insurmountable difficulties.

**Abbreviations**
LHVs  = local hidden variables
PBS  = polarization beam-splitter
QM  = quantum mechanics
SR  = special relativity

# 1. Introduction

Just three decades after the famous EPR article [1] which questioned the completeness of the quantum mechanics (QM), J. Bell proved the impossibility of a certain type of local hidden variables (LHVs) to explain the results of experiments with quantum entanglements [2]. LHVs of this type possess a main feature: they obey the *probabilistic calculus with positive probabilities smaller than 1*.

Soon after Bell's inequalities, additional inequalities based on the same principles were found, and they are known under the name "Bell-type" [3 − 5]. The incompatibility between these inequalities and the quantum mechanics was proved by the famous experiments of the Aspect group [6, 7], and later on, by experiments of different other groups [8 − 11]. Thus, it became clear that the class of LHVs examined by Bell, does not exist.

However, the fact that a calculus with *positive probabilities* cannot reproduce the QM predictions should not be surprising. Bell and his followers ignored the fact that QM deals with wave-packets, not with billiard balls.

Although the wave-like behavior of the quantum objects was already well known from Jönsson's multi-slit experiment with electrons [12], Bell and his followers ignored it.

In manipulating wave-packets, the nature adds up their *amplitudes* (interference phenomenon), e.g. $\boldsymbol{a}_1 + \boldsymbol{a}_2$, not their *intensities*, $|\boldsymbol{a}_1|^2 + |\boldsymbol{a}_2|^2$. The total probability of obtaining a recording in a detector illuminated by two wave-packets of amplitude $\boldsymbol{a}_1$, respectively $\boldsymbol{a}_2$, is $|\boldsymbol{a}_1 + \boldsymbol{a}_2|^2 = |\boldsymbol{a}_1|^2 + |\boldsymbol{a}_2|^2 + 2\mathcal{R}e[\boldsymbol{a}_1^*\boldsymbol{a}_2]$. That differs from the sum $|\boldsymbol{a}_1|^2 + |\boldsymbol{a}_2|^2$. So, what was ruled out by the violation of Bell-type inequalities was not the locality, but one particular model of hidden variables.

The present text presents a *plausible* model of non-deterministic hidden variables which can reproduce the QM results. The model is quite parallel to Bell's model of LHVs, however, since it deals with waves, it takes as a basic physical concept the concept of amplitude, not the concept of probability. The probability of a detection is equal to the absolute square of the amplitude, as said above.

It is shown that the nonlocality appears unavoidably in consequence of the fact that the amplitudes are not real and positive numbers, but have phases with respect to the vacuum. For instance, given two particles A and B and two wave-packets, $|a\rangle$ of A with amplitude $\boldsymbol{a}_A$, and $|b\rangle$ of B with amplitude $\boldsymbol{a}_B$, the model assigns to the possibility that both wave-packets produce recordings, the amplitude of probability $\boldsymbol{a}_A\boldsymbol{a}_B$. In our experiments it may happen that the system evolves in different ways, e.g. one way that produces the amplitudes $\boldsymbol{a}_A$, respectively $\boldsymbol{a}_B$, and another way that produces the amplitudes $\boldsymbol{a'}_A$, respectively $\boldsymbol{a'}_B$. The amplitude of probability of the joint detection is, according to the model, the sum of the contributions, i.e. $\boldsymbol{a}_A\boldsymbol{a}_B + \boldsymbol{a'}_A\boldsymbol{a'}_B$. In the particular case that $|\boldsymbol{a}_A\boldsymbol{a}_B| = |\boldsymbol{a'}_A\boldsymbol{a'}_B|$ and the phases of the two contributions differ by $\pi$, the total amplitude vanishes. So, the wave-packets $|a\rangle$ and $|b\rangle$ can't be detected together. That imposes on the system a correlation between the experimental outcomes, and the effect is independent of the distance between the components of the system.

In continuation, the section 2 justifies where from comes the idea of full/empty waves. Section 3 presents the rules of the model, and sections 4 and 5 apply it to the particular cases of the polarization singlet and Sciarrino's experiment, showing that it predicts the same predictions as QM. Section 6 contains discussions.

## 2. Why Full/Empty Waves?

Consider the experiment in the figure below. From a signal-idler pair of photons the idler photon is sent to a detector D. The signal photon is sent to a beam-splitter BS which produces a reflected wave-packet $|a\rangle$ and a transmitted wave-packet $|b\rangle$. On the path of the wave-packet $|a\rangle$ is placed a detector E. Thus, if a detection occurs in E, it occurs after the detection of the idler photon in D, and that, by any frame of coordinates traveling in the direction $x$ or $-x$.

Up to this point the experiment is the same as the one performed by the Aspect group [13]. The difference is that in our case *no detector is placed on the path b*.

The question we are trying to answer is what can be inferred about the wave-packet $|b\rangle$ from the response of the detector E after the detector D clicked, *if there is no detector on the path b*.

The QM gives us for the projection operators $|a\rangle\langle a|$ and $|b\rangle\langle b|$ the completion equality

$$|b\rangle\langle b| = 1 - |a\rangle\langle a|. \tag{1}$$

After the click in D the Hamiltonian describing the signal photon beyond BS, contains besides $\hat{a}^\dagger\hat{a} + \hat{b}^\dagger\hat{b}$, terms of the type $\hat{a}^\dagger\hat{b}$ and $\hat{b}^\dagger\hat{a}$, where $\hat{a}$ and $\hat{a}^\dagger$ are the lowering and raising operators for the photon on the path $a$, while $\hat{b}$ and $\hat{b}^\dagger$ are the similar operators for the photon on the path $b$. Therefore, it's obvious that the projection operator $|a\rangle\langle a|$ doesn't commute with the Hamiltonian, and neither does $|b\rangle\langle b|$. In other words, there is no reason to think that the number of photons on the paths is well determined before detection. The question is whether the detection on the path $a$ is sufficient for fixing the number of photons on the path $b$.

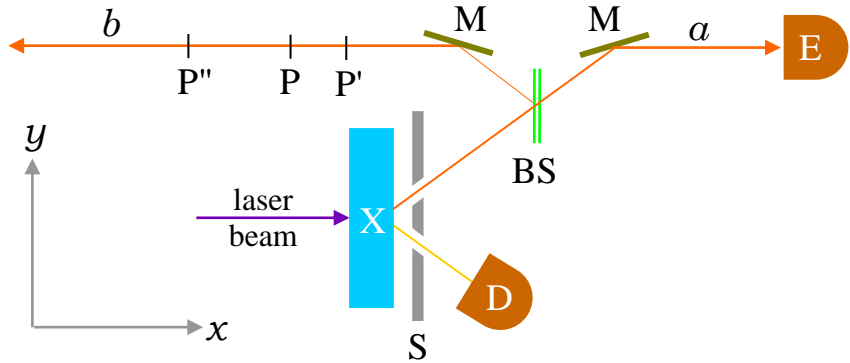

**Figure 1.** Single photon anti-bunching.
(The colors are only for clarity.) A laser UV beam illuminates a nonlinear crystal X producing signal-idler pairs. Two thin holes in the screen S select pairs from which the idler photon is redirected by means of a mirror M to the detector C, and the signal photon to the beam-splitter BS. See the rest of the explanations in the text.

Let's see if the special relativity (SR) theory can give some answer.
We discuss below a case in which both detectors D and E click, i.e. the projector $|a\rangle\langle a|$ gets the value 1. Let's denote by P the point on the path $b$ which is, by the lab frame, at the same distance from BS as the detector E.

**a)** Consider now a frame of coordinates F' traveling in the direction $x$. On the time axis of this frame, when the wave-packet $|a\rangle$ reaches the detector E the wave-packet $|b\rangle$ reaches a point P' closer to BS than P.
On the other hand, by the equality (1), for any point beyond P' (including P) the operator $|b\rangle\langle b|$ gets the value 0. The higher is the frame velocity $v$, the closer is the point P' to BS. So, one can infer in the limit $v \to c$ that the wave-packet $|b\rangle$ exited BS as an *empty wave*.

**b)** Consider a frame of coordinates F'' traveling in the direction $-x$. According to the time axis of this frame, by the time the wave-packet $|a\rangle$ reaches the detector E the wave-packet $|b\rangle$ reaches a point P'' more distant from BS than P. Therefore, for any point closer to BS than P'' (including P), the equality (1) can assign no defined value to the projector $|b\rangle\langle b|$.

The higher is the absolute velocity $|v|$ of the frame, the more distant is the point P'' from BS. So, one can infer that in the limit $|v| \to c$ *it is undefined* whether the wave-packet $|b\rangle$ is full or empty.

We got two different conclusions about the wave-packet $|b\rangle$. One can argue that there is no contradiction between them, i.e. that $|b\rangle$ is an empty wave though the frame F'' just doesn't allow to be aware of that.

Another experiment that makes the idea of full/empty waves appealing is the Elitzur-Vaidman proposal for nondestructive detection [14]. Imagine that the two paths $a$ and $b$ take the same direction and meet on a second beam-splitter, and they are of equal length, figure 2. If the two path are free, the signal photon will be detected in the detector C, never in the detector G. However, if an object B (a *bomb*, in the terminology of reference [14]) is placed on the path $b$, there is a probability of only 50% that the bomb won't explode and the photon would exit the interferometer, and 25% that the photon would be detected in G.

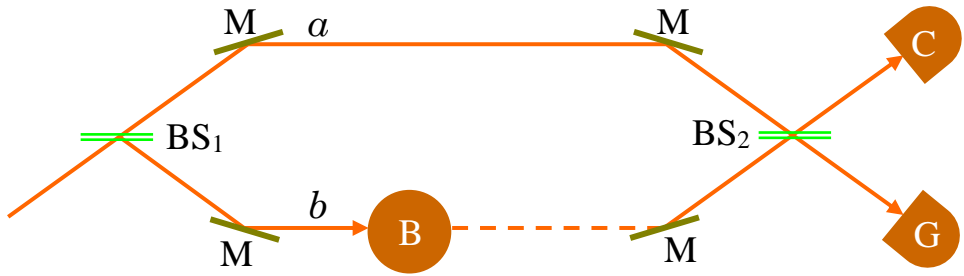

**Figure 2.** Nondestructive measurement.
See explanations in the text.

A detection in G testifies that the bomb is present in the interferometer even if it doesn't explode. The explanation for the signal photon reaching this detector, of course, is that the wave-packet $|b\rangle$ was stopped by the bomb. But, why the bomb didn't explode at the contact with this wave-packet? An appealing explanation is that the wave-packet is an empty wave.

In continuation, the sections $3 - 5$ bring additional arguments in favor of this idea, by presenting a model for entanglements based on it, and showing on a couple of known experiments how well the model reproduces the QM predictions. What is more, it will be shown that the model explains in a natural way the QM is nonlocal.

Unfortunately, the last section will show that this idea encounters also big difficulties.

## 3. A Full/Empty Wave Model for Entanglements

The model proposed below is for 2-particles, A and B, entangled by polarization, but it can be easily extended to other observables and to more particles. The model assumes the following:

**a)** For each pair of particles, an arbitrary direction $\vec{u}$ is picked in the polarization plane. We denote by $\vec{v}$ the direction orthogonal to $\vec{u}$ in this plane.

**b)** The source emits for each particle in the pair two wave-packets, one of polarization $\vec{u}$ and one of polarization $\vec{v}$. The wave-packet $|\vec{u}\rangle$ is defined as a *full* wave, while $|\vec{v}\rangle$ as an *empty* wave.

**c)** Empty waves and full waves behave exactly in the same way (except for their polarizations), up to the detector: the full wave impresses a detector while the empty wave doesn't.

**d)** The probability of a joint detection, e.g. particle A in a detector M and particle B in a detector N, is equal with the absolute square of the joint amplitude of the *combination of full waves* $|\vec{m}\rangle$ of A and $|\vec{n}\rangle$ of B.

**e)** The joint amplitude of two wave-packets, $|\vec{x}\rangle$ of A and $|\vec{y}\rangle$, is equal to the product of the separate amplitudes of the two wave-packets.

**f)** If there are a couple of contributions to a wave-packet, or to a combination of wave-packets, the resulting amplitude is equal to the sum of the amplitudes of the contributions.

**g)** When coming to a beam-splitter, the full wave property passes arbitrarily to one of the outputs, except for outputs which are blocked (see case 1 of section 4).

**h)** The full wave feature prevails over the empty wave feature. The meaning will become clear in section 4.

## 4. Applying the Model to the Polarization Singlet

For shortening the formulas we make the convention that in each combination (product) of single-particle wave-packets, we write on the left the wave-packet of the particle A and on the right the wave-packet of the particle B. To stress the full wave-packets we will write them in boldface.

Consider now a pair of particles A and B in the polarization singlet state.
According to the rule **a** of the model presented in the previous section, for a given pair of particles an arbitrary direction $\vec{u}$ is picked in the polarization plane. According to the rule **b**, we assume that the source emits for each particle in the pair, a wave-packet $|\vec{u}\rangle$ of amplitude $2^{-1/4}$ and a wave-packet $|\vec{v}\rangle$ of the same amplitude. Only the combinations $|\vec{u}\rangle|\vec{u}\rangle$ and $|\vec{v}\rangle|\vec{v}\rangle$ have nonzero joint amplitudes, equal to $2^{-1/2}$.

**Case 1.** Assume that the two particles' polarizations are tested in the same system of axes, $\{\vec{x}, \vec{y}\}$, which differs from the original (hidden) system of axes $\{\vec{u}, \vec{v}\}$ by a rotation of angle $\theta$ in the polarization plane. Then the wave-packets $|\vec{u}\rangle$ and $|\vec{v}\rangle$ transform at the polarization beam-splitters (PBSs) as follows

$$\frac{1}{\sqrt[4]{2}}|\vec{u}\rangle \to \frac{\cos\theta}{\sqrt[4]{2}}|\vec{x}\rangle + \frac{\sin\theta}{\sqrt[4]{2}}|\vec{y}\rangle, \quad \text{or} \quad \frac{1}{\sqrt[4]{2}}|\vec{u}\rangle \to \frac{\cos\theta}{\sqrt[4]{2}}|\vec{x}\rangle + \frac{\sin\theta}{\sqrt[4]{2}}|\vec{y}\rangle, \tag{2}$$

$$\frac{1}{\sqrt[4]{2}}|\vec{v}\rangle = -\frac{\sin\theta}{\sqrt[4]{2}}|\vec{x}\rangle + \frac{\cos\theta}{\sqrt[4]{2}}|\vec{y}\rangle. \tag{3}$$

The coefficient before each wave-packet represents its amplitude. As show the formulas (2) and according to the rule **g** of the model, either the $\vec{x}$-polarized or the $\vec{y}$-polarized wave-packet, may become full. Let's assume for the moment the first option for the particle A, and the second for the particle B. Then, according to the rule **e**, the initial combination $|\vec{u}\rangle|\vec{u}\rangle$ produces the following joint amplitudes for wave-packet combinations:

### Table I

| Combination: | $\left|\vec{x}\right\rangle\left|\vec{x}\right\rangle$ | $\left|\vec{y}\right\rangle\left|\vec{y}\right\rangle$ | $\left|\vec{x}\right\rangle\left|\vec{y}\right\rangle$ | $\left|\vec{y}\right\rangle\left|\vec{x}\right\rangle$ |
|---|---|---|---|---|
| Joint amplitude: | $\dfrac{\cos^2\theta}{\sqrt{2}}$ | $\dfrac{\sin^2\theta}{\sqrt{2}}$ | $\dfrac{\sin\theta\cos\theta}{\sqrt{2}}$ | $\dfrac{\sin\theta\cos\theta}{\sqrt{2}}$ |

while the initial combination $\left|\vec{v}\right\rangle\left|\vec{v}\right\rangle$ produces

### Table II

| Combination: | $\left|\vec{x}\right\rangle\left|\vec{x}\right\rangle$ | $\left|\vec{y}\right\rangle\left|\vec{y}\right\rangle$ | $\left|\vec{x}\right\rangle\left|\vec{y}\right\rangle$ | $\left|\vec{y}\right\rangle\left|\vec{x}\right\rangle$ |
|---|---|---|---|---|
| Joint amplitude: | $\dfrac{\sin^2\theta}{\sqrt{2}}$ | $\dfrac{\cos^2\theta}{\sqrt{2}}$ | $-\dfrac{\sin\theta\cos\theta}{\sqrt{2}}$ | $-\dfrac{\sin\theta\cos\theta}{\sqrt{2}}$ |

Finally, by to the rules **f** and **h**, it's obvious that only the combinations with the same polarization survive,

### Table III

| Combination: | $\left|\vec{x}\right\rangle\left|\vec{x}\right\rangle$ | $\left|\vec{y}\right\rangle\left|\vec{y}\right\rangle$ |
|---|---|---|
| Joint amplitude: | $\dfrac{1}{\sqrt{2}}$ | $\dfrac{1}{\sqrt{2}}$ |

Though, none of these results is possible. In each combination one of the wave-packets is empty i.e. for one of the particles the full wave property disappears. The rule **g** says that the things go otherwise, and here appears the nonlocal feature of this model. The decision on which wave-packet of a particle may become full cannot fall locally. The process evolves by trial and error: *the options that the full wave property of the particle A exit the PBS through the output port $\vec{x}$, and for the particle B through the output port $\vec{y}$, or vice-versa, are blocked.* The full wave of each particle would have to try another output until a combination of outputs of the same polarization is tried. Only then the full waves will succeed to exit the PBSs.

**Case 2.** A more complicated situation arises when the experimenters measure by different pairs of axes. Let these axes be $\{\vec{x},\vec{y}\}$ for the particle A, and $\{\vec{x}',\vec{y}'\}$ for Bob. The relationship between the axes $\{\vec{u},\vec{v}\}$ and $\{\vec{x},\vec{y}\}$ is a rotation by an angle $\theta$, while between the axes $\{\vec{u},\vec{v}\}$ and $\{\vec{x}',\vec{y}'\}$ is a rotation by an angle $\theta'$. Therefore, instead of the table I one gets,

### Table IV

| Combination: | $\left|\vec{x}\right\rangle\left|\vec{x}\right\rangle$ | $\left|\vec{y}\right\rangle\left|\vec{y}\right\rangle$ | $\left|\vec{x}\right\rangle\left|\vec{y}\right\rangle$ | $\left|\vec{y}\right\rangle\left|\vec{x}\right\rangle$ |
|---|---|---|---|---|
| Joint amplitude: | $\dfrac{\cos\theta\cos\theta'}{\sqrt{2}}$ | $\dfrac{\sin\theta\sin\theta'}{\sqrt{2}}$ | $\dfrac{\cos\theta\sin\theta'}{\sqrt{2}}$ | $\dfrac{\sin\theta\cos\theta'}{\sqrt{2}}$ |

and instead of table II,

#### Table V

| Combination: | $\lvert \vec{x} \rangle \lvert \vec{x} \rangle$ | $\lvert \vec{y} \rangle \lvert \vec{y} \rangle$ | $\lvert \vec{x} \rangle \lvert \vec{y} \rangle$ | $\lvert \vec{y} \rangle \lvert \vec{x} \rangle$ |
|---|---|---|---|---|
| Joint amplitude: | $\dfrac{\sin\theta\sin\theta'}{\sqrt{2}}$ | $\dfrac{\cos\theta\cos\theta'}{\sqrt{2}}$ | $-\dfrac{\sin\theta\cos\theta'}{\sqrt{2}}$ | $-\dfrac{\cos\theta\sin\theta'}{\sqrt{2}}$ |

Then, according to the rules **f** and **h**, there results

#### Table VI

| Combination: | $\lvert \vec{\boldsymbol{x}} \rangle \lvert \vec{x} \rangle$ | $\lvert \vec{y} \rangle \lvert \vec{\boldsymbol{y}} \rangle$ | $\lvert \vec{\boldsymbol{x}} \rangle \lvert \vec{\boldsymbol{y}} \rangle$ | $\lvert \vec{y} \rangle \lvert \vec{x} \rangle$ |
|---|---|---|---|---|
| Joint amplitude: | $\dfrac{\cos(\theta-\theta')}{\sqrt{2}}$ | $\dfrac{\cos(\theta-\theta')}{\sqrt{2}}$ | $-\dfrac{\sin(\theta-\theta')}{\sqrt{2}}$ | $\dfrac{\sin(\theta-\theta')}{\sqrt{2}}$ |

Thus, in the case 2 the restriction of same polarization for full waves disappears. One can also see that in both cases 1 and 2, the hidden initial axes $\vec{u}$ and $\vec{v}$ play no role in the final results.

It's easy to compare the final results of the two cases with the QM predictions and see that they are the same.

## 5. Applying the Model to Sciarrino's Experiment

The figure 3 shows the schema of an experiment similar to the one performed by F. Sciarrino [16]. Pairs of photons are produced by a nonlinear crystal X through *degenerate* down-conversion of UV photons. Each one of the photons in the pair, lands on a PBS suitably oriented s.t. from each PBS exit a $u$-polarized wave-packet in the direction of Eve and a $v$-polarized wave-packet in the direction of Victor. All the wave-packets are of the same intensity. At Eve's site the wave-packets from the two particles meat on a beam-splitter BS beyond which there are two detectors, C and D. At Victor's site the two wave-packets intersect, and a recording system R reports the position at which the photon was detected. All the paths from the PBSs to the detectors C, D, and R, are equal, and only single-photon detections in coincidence between C and R, or D and R, are reported.

The effect is that for the photons detected at Eve's site in the same detector, e.g. in C, the recordings in R form an interference pattern.

Indeed, the wave-function of the pair is

$$\lvert\psi\rangle = \frac{\lvert u_1 \rangle + \lvert v_1 \rangle}{\sqrt{2}}\,\frac{\lvert u_2 \rangle + \lvert v_2 \rangle}{\sqrt{2}}. \tag{4}$$

At the beam-splitter BS occurs the transformation

$$\lvert u_1 \rangle \to \frac{1}{\sqrt{2}}\big(\iota\lvert c \rangle + \lvert d \rangle\big), \quad \lvert u_2 \rangle \to \frac{1}{\sqrt{2}}\big(\lvert c \rangle + \iota\lvert d \rangle\big), \tag{5}$$

therefore beyond BS the state of the system becomes

$$\lvert\psi\rangle = \frac{1}{2\sqrt{2}}\Big\{\iota\big(\lvert 2c \rangle + \lvert 2d \rangle\big) - \sqrt{2}\lvert v_1 \rangle\lvert v_2 \rangle + \big(\lvert v_1 \rangle - \iota\lvert v_2 \rangle\big)\lvert c \rangle + \iota\big(\lvert v_1 \rangle + \iota\lvert v_2 \rangle\big)\lvert d \rangle\Big\}. \tag{6}$$

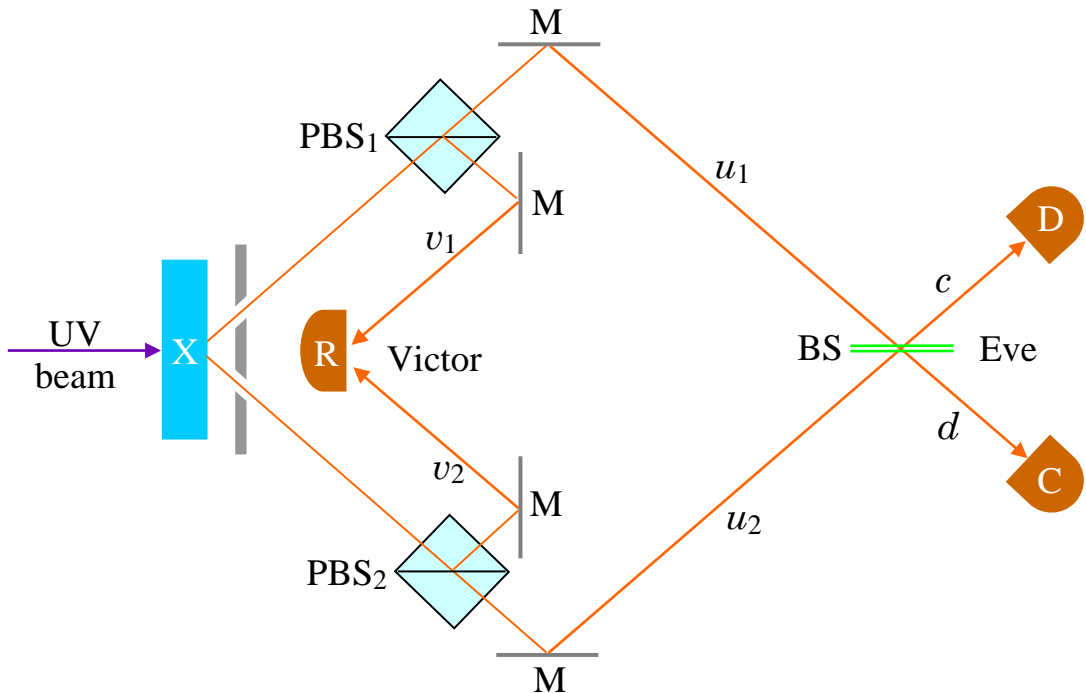

**Figure 3.** A Sciarrino-type experiment.
(The colors are only for clarity.) The explanations are given in the text.

In the cases $|2c\rangle$, $|2d\rangle$, and $|v_1\rangle|v_2\rangle$ there is no coincidence between the detectors of the two sites, s.t. these cases are not recorded. However, one can see from the wave-function (6) that if the detections in R coincide with detections in C, they form an interference pattern. (Similarly, for the detections in R coinciding with detections in D.)

These results can be explained in the terminology of the *full/empty* wave model from section 3, as follows: a detection in the detector D may be the result of $|u_1\rangle$ ($|u_2\rangle$) being a full wave that passed to $|d\rangle$. Then, $|v_1\rangle$ ($|v_2\rangle$) is an empty wave, and according with the rules **c** and **f** of the model, what is recorded in R is the interference tableau produced by the full wave $|v_2\rangle$ ($|v_1\rangle$) with the empty wave $|v_1\rangle$ ($|v_2\rangle$).

An interesting conclusion about the empty waves – if such things exist in reality – is that in which-way experiments as in section 2, these waves can't be detected, however in interference experiments, as the one described in the present section, they prove their presence.

## 6. Discussions – an impasse!

The model proposed above goes, up to a certain point, parallel with the LHV model used by Bell: a hidden direction of polarization, $\bar{u}$, arbitrarily picked, is assigned to full waves. Together with that, empty waves are emitted. The rules of the calculus are similar with those used by Bell, except that they act on amplitudes instead of probabilities: each wave-packet has an amplitude, which is a local quantity. Joint measurements are related with joint amplitudes of probability obtained as products of the amplitudes of the wave-packets. If there are several contributions, the joint amplitudes are added, instead of the joint probabilities.

Probabilities are not a basic concept, they are obtained from amplitudes by taking the absolute square.

Here is the root of the *nonlocality* of the QM. Joint amplitudes are complex numbers, not real and positive. Therefore, different contributions to the joint amplitude of a combination of results, may cancel one another s.t. the combination is simply *erased*, as happened in the case 1 of section 4. The full wave property of the particles may exit the PBSs only if an allowed combination of full waves is formed.

The model seems quite appealing, and seems to be in full agreement with the QM predictions for entanglements.

Though, A. Suarez brought the objection that the idea of full/empty waves entails the many worlds assumption [17]. The rationale goes as follows: if a full wave $|a\rangle$ and an empty wave $|b\rangle$ exit a beam-splitter BS, then *by symmetry*, also a full wave $|b\rangle$ and an empty wave $|a\rangle$ should exit BS at the same time. The full wave $|b\rangle$ and the empty wave $|a\rangle$ are bound to be tested and found as such, in *another world*, with *other experimenters* living in it. Suarez' objection holds as long as there is no reason for symmetry-breaking at the beam-splitter. Though, it can be escaped by invoking the non-determinism of the quantum world. Indeed, one can reply that in each trial and trial of the experiment, the full wave is picked at random but the symmetry is achieved by that in half of the trials the full wave property goes to one output of the BS, and in the other half to the other output.

Another difficulty with the full/empty wave assumption arises from an unavoidable requirement: the full wave property has to follow a continuous path. Indeed, this property cannot jump from a wave-packet traveling one path to a wave-packet traveling another path, otherwise, a frame of coordinates can be found in which this property appears at once on both path. In that case the particle could be detected at once on both paths, thereby violating the energy conservation.

This implication entails additional consequences, which speak against the assumption itself, as pointed out by the present author [18]. The analysis in [18] relies on the so-called "Hardy's paradox" [19], which examined the evolution of a system of two particles from the point of view of different observers in relative movement. For a certain combination of results of the experiment suggested by Hardy, each observer infers about one of the particles that in the past, it should have been *with certainty* on a certain path. In the full/empty waves terminology, that means that the full wave of that particle should have been on that path. But the resulting pair of paths for the two particles, is forbidden. That wouldn't be a problem if the continuous path requirement wouldn't interdict a full wave to jump from one path to another. So, we come to a *contradiction*.

There are two waves to escape the contradiction: one is admit that the reasoning according to moving frames is wrong, in other words, only the conclusions obtained according to one frame – a preferred frame – are correct. But according to SR, all the frames are valid. Experiments were performed either with beam-splitters in relative movement or with detectors in relative movements [20 – 22], but no violation of the wave-function in some frame, was observed.

Another escape would be to abandon the full/empty wave hypothesis. In other words, results of measurements are not pre-determined, they fall at the measurement time. Though, the situation remaining in this case is no easier: the model described in section 3 becomes invalid and without it, it's much more complicated to explain the mechanism of the quantum correlations.

But an even worse consequence appears. According to the SR, measurements of entangled particles which are simultaneous according to some frame of coordinates F, by another frame F' may be not simultaneous. Judging according to the frame F', results of measurement done *now* are interdependent with results which will be obtained in the *future*, and which are not predetermined by any realities existing *now*. A. Suarez states in different articles and lectures that quantum entanglements live outside the space-time [23 – 25], or, to quote a recent expression of him, measurement results of quantum entanglements are "*not determined by any properties pre-existing in space-time*". However, these statements don't explain much, they don't offer an exit from the absurd scenario that the future may influence the present.

The only conclusion is that for solving this impasse additional experiments are needed.

## Acknowledgements

I am deeply thankful to Prof. A. Suarez from the Center for Quantum Philosophy, Zurich, for the profound dialogue on different problems regarding quantum entanglements and the full/empty wave hypothesis.